\pdfoutput=1
\documentclass[12pt]{article}

\usepackage[margin=1in]{geometry}
\usepackage[T1]{fontenc}
\usepackage{amsmath,amssymb}
\usepackage{graphicx}
\usepackage{booktabs}
\usepackage{array}
\usepackage{caption}
\usepackage{subcaption}
\usepackage{float}
\usepackage{microtype}
\usepackage{url}
\usepackage{setspace}
\usepackage{authblk}
\usepackage[numbers]{natbib}
\usepackage[hidelinks]{hyperref}

\title{\textbf{When Method Choice Changes Statistical Inference:\\
A Comparison of a Baseline Two-Stage Approach and Bayesian Joint
Modeling for Longitudinal and Survival Data in an HIV Clinical Trial}}

\author[1]{Alberta A. Johnson\thanks{Corresponding author: \texttt{john1269@bears.unco.edu}}}
\affil[1]{Department of Applied Statistics and Research Methods, University of Northern Colorado, Greeley, Colorado, USA}
\date{}

\begin{document}

\maketitle

% ============================================================
% ABSTRACT
% ============================================================
\begin{abstract}
The two-stage approach and Bayesian joint modeling are commonly used to
analyze longitudinal biomarker measurements together with time-to-event
outcomes. Using data from an HIV clinical trial of 467 patients with
repeated CD4 measurements and all-cause mortality as the survival
outcome, we compared a baseline two-stage approach with a Bayesian joint
model. The two-stage analysis fitted a linear mixed-effects model and
included each patient's predicted baseline CD4 value as a fixed
covariate in a Cox proportional hazards model. The joint model
simultaneously modeled the longitudinal CD4 process and survival while
linking mortality risk to the current underlying CD4 value. The
estimated association between ddI and mortality was similar in direction
and magnitude across the two approaches. The two-stage estimate was
HR $=1.342$ (95\% CI: $1.006$--$1.789$), whereas the joint-model estimate
was HR $=1.383$ (95\% CrI: $0.952$--$2.010$). The estimated protective
association of CD4 was stronger under the joint model (HR $=0.776$) than
under the two-stage approach (HR $=0.826$). Because the approaches used
different summaries of the longitudinal CD4 process, the observed
differences cannot be attributed solely to measurement error or
informative dropout. The findings demonstrate that the treatment of
longitudinal biomarker information can materially affect statistical
inference.
\end{abstract}

\noindent\textbf{Keywords:} joint models; longitudinal data; survival
analysis; Bayesian estimation; two-stage approach; HIV; CD4 cell count;
MCMC

% ============================================================
\section{Introduction}
% ============================================================

In clinical and epidemiological research, repeated measurements of a
biomarker are often collected together with a time-to-event outcome for
the same individuals. Studies of HIV progression, cancer treatment, and
chronic disease management routinely generate this type of joint
longitudinal-survival data. Appropriate analysis is important because
the biomarker process and the event process may be related, and
inadequate treatment of that relationship can lead to biased or
misleading estimates.

A common applied strategy is the two-stage approach. In the first stage,
a linear mixed-effects model is fitted to the longitudinal biomarker
data to obtain patient-specific predicted values. In the second stage,
those predictions are included as covariates in a Cox proportional
hazards model for the survival outcome.\cite{Cox1972} Although intuitive
and relatively easy to implement, this strategy has well-documented
limitations.\cite{Prentice1982} In particular, it may treat estimated
biomarker values as fixed and known, thereby failing to propagate
uncertainty from the longitudinal model into the survival model. It may
also inadequately address informative dropout when patients who
experience the event earlier contribute fewer longitudinal
measurements.\cite{Tsiatis1995}

Joint models address these issues by modeling the longitudinal and
survival processes within a unified framework.\cite{Wulfsohn1997} The
longitudinal submodel characterizes the underlying biomarker trajectory,
while the survival submodel links the instantaneous event hazard to a
feature of that trajectory, such as its current value. Under appropriate
model specification, this shared-parameter framework can account for
measurement error in the biomarker process and dependence between
longitudinal follow-up and the event process.\cite{Wulfsohn1997,Tsiatis2004}

The theoretical basis for joint modeling is well established. Tsiatis
et al. demonstrated that naive incorporation of longitudinal covariates
into survival models can produce biased estimates when the covariate is
measured with error.\cite{Tsiatis1995} Wulfsohn and Tsiatis proposed a
foundational joint modeling framework that estimates the longitudinal
and survival processes simultaneously.\cite{Wulfsohn1997} Rizopoulos
later provided a comprehensive treatment of joint models and accessible
software implementations in R.\cite{Rizopoulos2012}

Despite these developments, two-stage analyses remain common in applied
work because they are familiar, easy to implement, and computationally
convenient. Previous studies have compared two-stage and joint-modeling
strategies theoretically and through simulation,\cite{Prentice1982,Tsiatis1995}
but practical comparisons using accessible clinical datasets remain
useful for illustrating how modeling choices affect interpretation.

This paper uses data from a randomized clinical trial comparing
didanosine (ddI) and zalcitabine (ddC) in HIV-positive
patients.\cite{Goldman1996} We compare a baseline two-stage approach
with a Bayesian joint model for the relationship between repeated CD4
measurements and time to death. The purpose is not to claim that one
method alone changes the underlying clinical effect. Rather, the study
illustrates how alternative representations of longitudinal biomarker
information can produce different inferential statements. The Bayesian
joint model was fitted using the freely available \texttt{JMbayes2}
package in R.\cite{Rizopoulos2023}

The remainder of the paper is organized as follows. Section~2 describes
the data and statistical methods. Section~3 presents the results.
Section~4 discusses their implications and limitations. Section~5
concludes.

% ============================================================
\section{Methods}
% ============================================================

\subsection{Data}

We used the AIDS clinical trial dataset available in the
\texttt{JMbayes2} package in R,\cite{Rizopoulos2023} originally derived
from a randomized controlled trial comparing didanosine (ddI) and
zalcitabine (ddC) in HIV-positive patients.\cite{Goldman1996} The
dataset included 467 patients followed for up to 21.4 months. The
longitudinal outcome was the square root of CD4 cell count, measured at
baseline and approximately every 6 months, yielding 1,405 observations.
The number of CD4 measurements per patient ranged from 1 to 5,
reflecting unbalanced follow-up due to dropout and early death. The
survival outcome was time to death, with 188 deaths observed (40.3\%
event rate). Baseline covariates included drug assignment (ddI vs.\ ddC),
prior opportunistic infection (AIDS vs.\ no AIDS), and prior AZT use
(failure vs.\ intolerance).

\subsection{Statistical Analysis}

All analyses were conducted in R version 4.5 using \texttt{nlme} version
3.1-164 and \texttt{JMbayes2} version 0.4-5.\cite{RCore2025,Pinheiro2023,Rizopoulos2023}

\subsubsection{Baseline Two-Stage Approach}

In the first stage, we fitted a linear mixed-effects model for the
square-root-transformed CD4 trajectory. Fixed effects were included for
observation time, drug assignment, prior opportunistic infection, and
prior AZT use. Random intercepts and random slopes for time were included
at the patient level:

\begin{equation}
Y_{ij}
=
\beta_0
+
\beta_1 t_{ij}
+
\beta_2 \text{drug}_i
+
\beta_3 \text{prevOI}_i
+
\beta_4 \text{AZT}_i
+
b_{0i}
+
b_{1i}t_{ij}
+
\varepsilon_{ij},
\label{eq:lme}
\end{equation}

\noindent where $Y_{ij}$ is the square root of the CD4 count for patient
$i$ at time $t_{ij}$,
$\mathbf{b}_i=(b_{0i},b_{1i})^\top
\sim \mathcal{N}(\mathbf{0},\mathbf{D})$ are patient-specific random
effects, and
$\varepsilon_{ij}\sim\mathcal{N}(0,\sigma^2)$ is the residual error.

In the second stage, a patient-specific predicted CD4 value was
extracted at baseline and entered as a time-fixed covariate in a Cox
proportional hazards model together with drug assignment, prior
opportunistic infection, and prior AZT use:

\begin{equation}
h(t\mid\mathbf{X}_i)
=
h_0(t)
\exp\left(
\gamma_1\widehat{m}_i
+
\gamma_2\text{drug}_i
+
\gamma_3\text{prevOI}_i
+
\gamma_4\text{AZT}_i
\right),
\label{eq:cox}
\end{equation}

\noindent where $\widehat{m}_i$ is the predicted baseline square-root
CD4 value for patient $i$ from Stage~1 and $h_0(t)$ is the unspecified
baseline hazard.

This baseline two-stage approach treats the predicted baseline CD4 value
as fixed and does not propagate uncertainty from the longitudinal model
into the Cox model. It also differs from the joint model because it uses
a single baseline summary rather than the evolving CD4 trajectory.

\subsubsection{Bayesian Joint Model}

We fitted a Bayesian joint model using the \texttt{JMbayes2}
package.\cite{Rizopoulos2023} The model simultaneously estimated a
linear mixed-effects submodel for CD4 and a survival submodel for time to
death. The longitudinal submodel had the same form as
Equation~\eqref{eq:lme}. The survival submodel was

\begin{equation}
h(t\mid\mathbf{X}_i,\mathbf{b}_i)
=
h_0(t)
\exp\left(
\alpha m_i(t)
+
\gamma_2\text{drug}_i
+
\gamma_3\text{prevOI}_i
+
\gamma_4\text{AZT}_i
\right),
\label{eq:joint}
\end{equation}

\noindent where

\begin{equation}
m_i(t)
=
\beta_0
+
\beta_1 t
+
\beta_2\text{drug}_i
+
\beta_3\text{prevOI}_i
+
\beta_4\text{AZT}_i
+
b_{0i}
+
b_{1i}t
\end{equation}

\noindent is the current underlying square-root CD4 trajectory and
$\alpha$ is the association parameter linking the longitudinal and
survival submodels. The joint model therefore relates mortality risk to
the current underlying CD4 value rather than to a single predicted
baseline value.

The two approaches do not use identical representations of the
longitudinal exposure. Their estimates should therefore be interpreted
as a practical comparison of two analysis strategies rather than as an
isolated test of measurement-error correction.

The model was estimated using Markov chain Monte Carlo sampling with
three parallel chains, 12,000 iterations per chain, 2,000 burn-in
iterations, and a thinning factor of 5. This yielded 2,000 retained
posterior samples per chain. The default prior and baseline-hazard
settings in \texttt{JMbayes2} version 0.4-5 were retained. Convergence
was assessed visually using traceplots and by Gelman--Rubin statistics,
with all $\widehat{R}<1.01$.\cite{Gelman1992}

\subsection{Model Comparison}

The approaches were compared with respect to four quantities:
(1) the estimated association between CD4 and mortality,
(2) the association between drug assignment and mortality,
(3) the association between prior opportunistic infection and mortality,
and
(4) the association between prior AZT use and mortality.

For the two-stage approach, we report hazard ratios (HRs), 95\%
confidence intervals (CIs), and Wald-test $p$-values. For the Bayesian
joint model, we report posterior HRs and 95\% credible intervals (CrIs).
Because the two-stage model used predicted baseline CD4 and the joint
model used the current underlying CD4 value, the corresponding CD4 HRs
do not have identical interpretations.

% ============================================================
\section{Results}
% ============================================================

\subsection{Data Description}

The dataset comprised 467 HIV-positive patients followed for a median of
16.97 months (range: 0.5 to 21.4 months). A total of 1,405 CD4
measurements were recorded, with patients contributing between 1 and 5
observations. The overall event rate was 40.3\%, with 188 deaths observed
during follow-up. The study included 237 patients randomized to ddC and
230 randomized to ddI.

\begin{table}[H]
\centering
\caption{Descriptive summary of the AIDS clinical trial dataset
($N=467$).}
\label{tab:descriptive}
\begin{tabular}{lcc}
\toprule
\textbf{Variable} & \textbf{Category/Statistic} & \textbf{Value} \\
\midrule
\multicolumn{3}{l}{\textit{Survival outcome}} \\
\quad Follow-up time (months) & Median (range) & 16.97 (0.5--21.4) \\
\quad Deaths & $n$ (\%) & 188 (40.3\%) \\
\midrule
\multicolumn{3}{l}{\textit{Longitudinal outcome}} \\
\quad CD4 measurements & Total & 1,405 \\
\quad Measurements per patient & Range & 1--5 \\
\quad CD4 count, square-root scale & Mean (SD) & 7.02 (3.91) \\
\midrule
\multicolumn{3}{l}{\textit{Baseline covariates}} \\
\quad Drug assignment & ddC & 237 (50.7\%) \\
& ddI & 230 (49.3\%) \\
\quad Prior opportunistic infection & AIDS & 232 (49.7\%) \\
& No AIDS & 235 (50.3\%) \\
\quad Prior AZT use & Intolerance & 241 (51.6\%) \\
& Failure & 226 (48.4\%) \\
\bottomrule
\end{tabular}
\end{table}

\subsection{Descriptive Analysis}

\begin{figure}[H]
\centering
\includegraphics[width=0.6\textwidth]{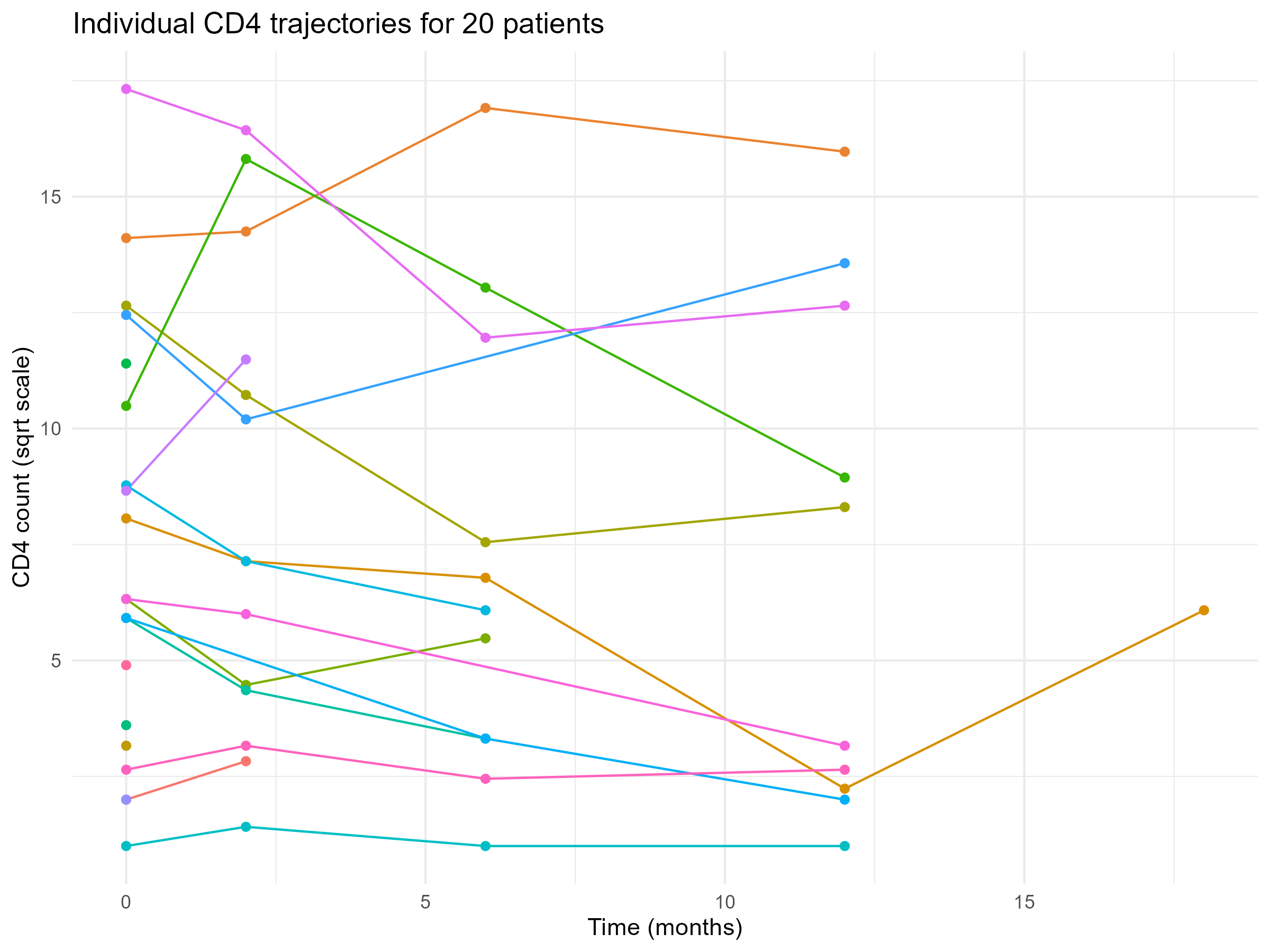}
\caption{Individual CD4 trajectories on the square-root scale for 20
randomly selected patients.}
\label{fig:individual}
\end{figure}

Individual CD4 trajectories revealed substantial between-patient
variability in both baseline levels and rates of change over time
(Figure~\ref{fig:individual}). Although the overall pattern suggested a
decline in CD4 during follow-up, some patients showed stable or
increasing trajectories, supporting the use of patient-specific random
effects.

\begin{figure}[H]
\centering
\includegraphics[width=0.6\textwidth]{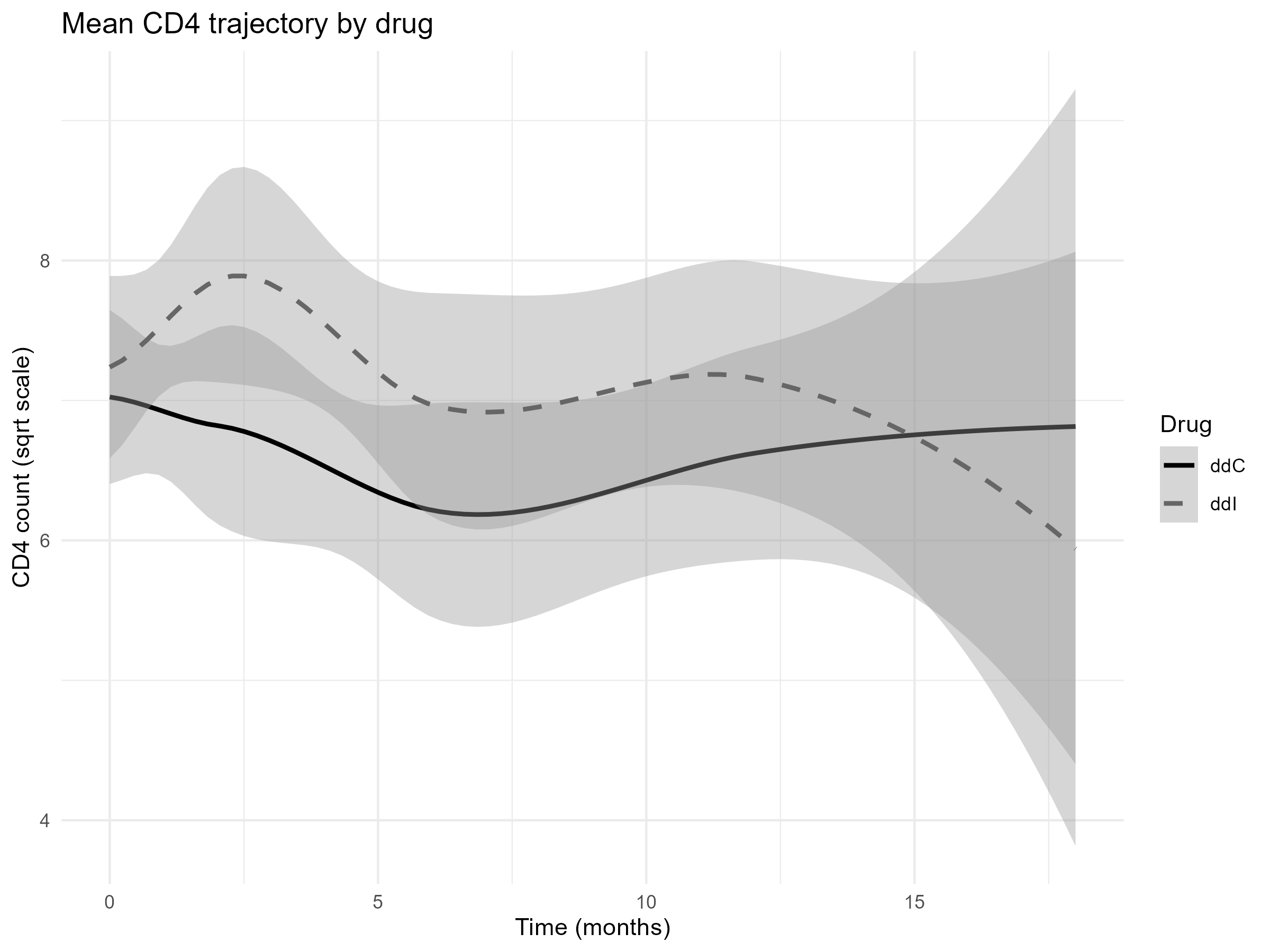}
\caption{Mean CD4 trajectory on the square-root scale by drug assignment,
estimated using LOESS.}
\label{fig:mean}
\end{figure}

The descriptive mean CD4 trajectories by drug group are shown in
Figure~\ref{fig:mean}. The LOESS curves suggested that the groups may
have differed in their average patterns over time. However, these curves
are descriptive, and the fitted longitudinal model did not include a
time-by-drug interaction.

\begin{figure}[H]
\centering
\includegraphics[width=0.65\textwidth]{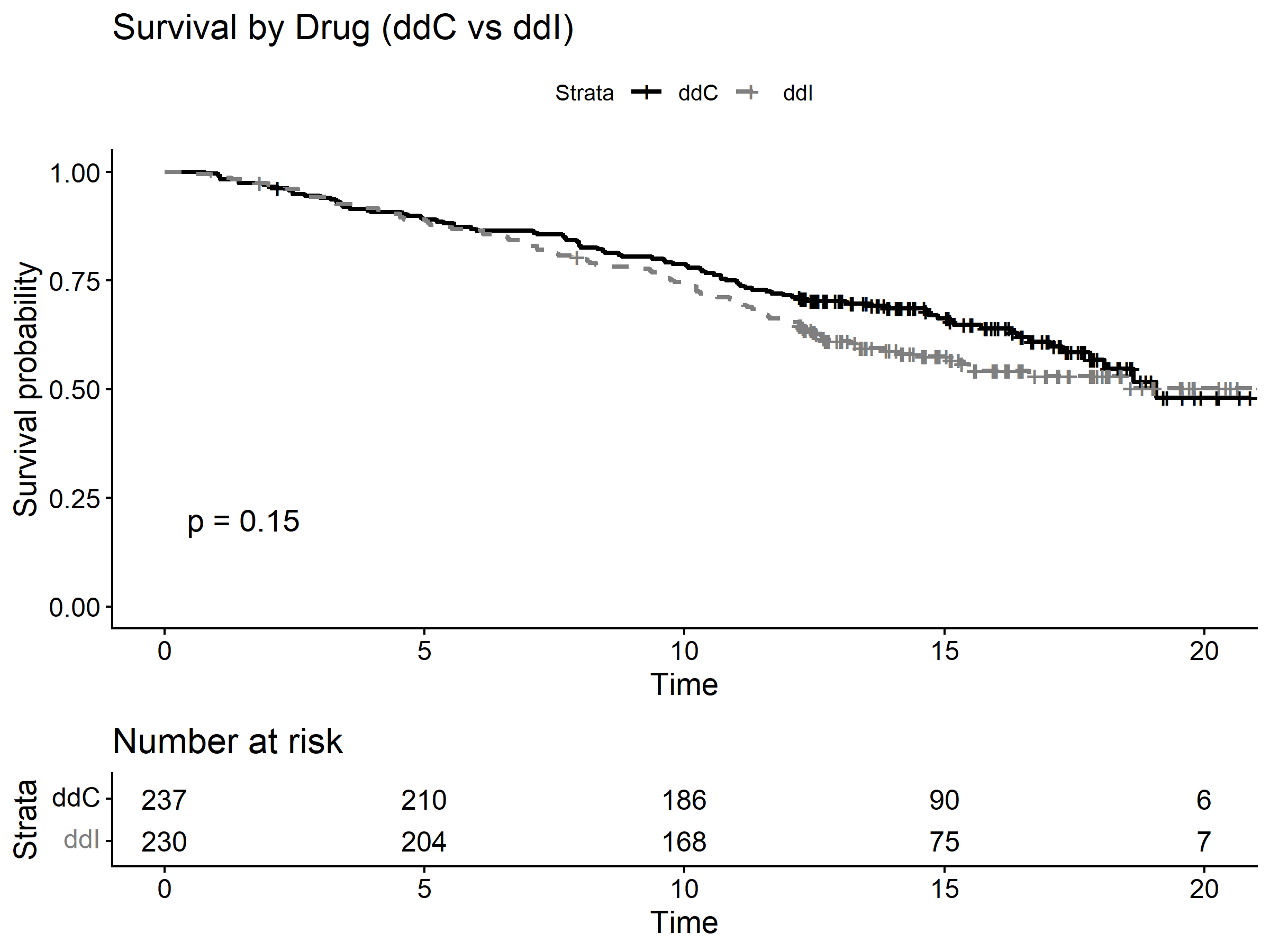}
\caption{Kaplan--Meier survival curves by drug assignment
(log-rank $p=0.15$).}
\label{fig:km}
\end{figure}

Kaplan--Meier survival curves are shown in Figure~\ref{fig:km}. The
unadjusted log-rank test did not identify a statistically significant
difference between the drug groups ($p=0.15$).

\subsection{Baseline Two-Stage Approach}

In Stage~1, square-root CD4 declined over time
($\widehat{\beta}_1=-0.153$ per month, $p<0.001$). Prior opportunistic
infection was associated with a lower baseline square-root CD4 value
($\widehat{\beta}_3=-4.628$, $p<0.001$). Drug assignment was not
associated with baseline square-root CD4 ($p=0.234$), and prior AZT use
was also not associated with baseline square-root CD4 ($p=0.562$).
Because the model did not include a time-by-drug interaction, it assumed
the same average rate of CD4 change for both treatment groups.

In Stage~2, a higher predicted baseline square-root CD4 value was
associated with a lower hazard of death
(HR $=0.826$, 95\% CI: $0.781$--$0.873$, $p<0.001$).
After adjustment for predicted baseline CD4 and the other covariates,
ddI was associated with a higher hazard of death than ddC
(HR $=1.342$, 95\% CI: $1.006$--$1.789$, $p=0.045$).
Prior opportunistic infection was associated with a higher hazard of
death
(HR $=1.910$, 95\% CI: $1.203$--$3.031$, $p=0.006$).
Prior AZT failure was not statistically significant
(HR $=1.111$, 95\% CI: $0.810$--$1.522$, $p=0.515$).
The model concordance statistic was 0.724.

\subsection{Bayesian Joint Model}

The Bayesian joint model showed adequate convergence across all three
MCMC chains. Gelman--Rubin statistics were below 1.01 for all monitored
parameters, and visual inspection of the traceplots in
Figure~\ref{fig:traceplots} indicated good mixing and stationarity.

The longitudinal submodel estimated a decline in square-root CD4 over
time
(posterior mean $=-0.179$, 95\% CrI: $-0.210$ to $-0.148$).
Prior opportunistic infection was associated with a lower baseline
square-root CD4 value
(posterior mean $=-4.697$, 95\% CrI: $-5.650$ to $-3.763$).

In the survival submodel, a higher current underlying square-root CD4
value was associated with a lower hazard of death. The posterior
log-hazard coefficient was $-0.254$
(95\% CrI: $-0.328$ to $-0.186$), corresponding to an HR of 0.776
(95\% CrI: $0.720$--$0.830$). Thus, each one-unit increase in the
current underlying square-root CD4 value was associated with an
estimated 22.4\% lower hazard of death.

The posterior HR for ddI versus ddC was 1.383
(95\% CrI: $0.952$--$2.010$). The posterior estimate suggested a higher
hazard for ddI, although the credible interval included 1. Prior
opportunistic infection was associated with a higher hazard of death
(HR $=1.801$, 95\% CrI: $1.103$--$2.999$). The credible interval for
prior AZT failure included 1
(HR $=1.089$, 95\% CrI: $0.781$--$1.523$).

\begin{figure}[p]
\centering
\begin{subfigure}[b]{0.66\textwidth}
\centering
\includegraphics[width=\textwidth]{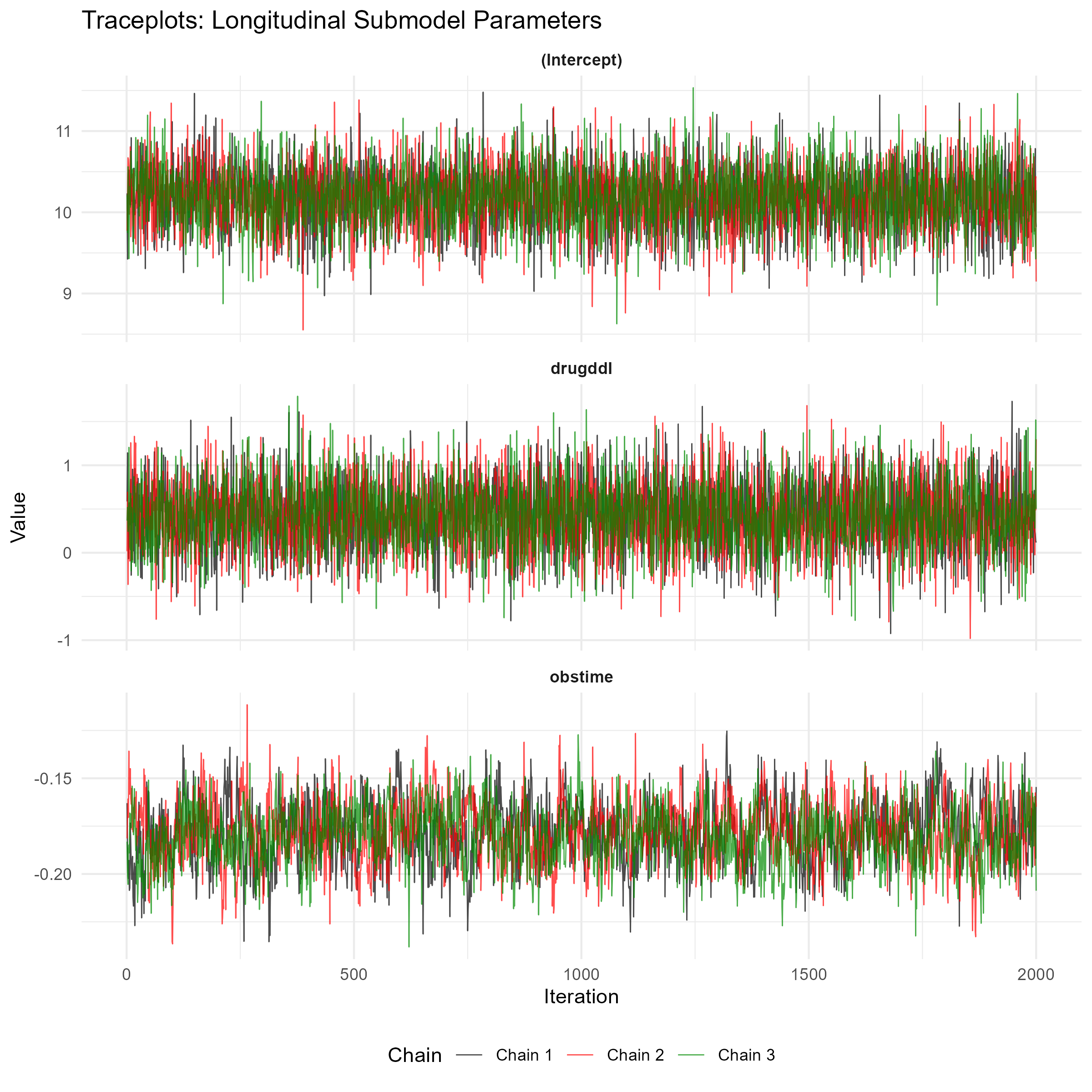}
\caption{Longitudinal submodel parameters.}
\end{subfigure}

\vspace{0.35cm}

\begin{subfigure}[b]{0.66\textwidth}
\centering
\includegraphics[width=\textwidth]{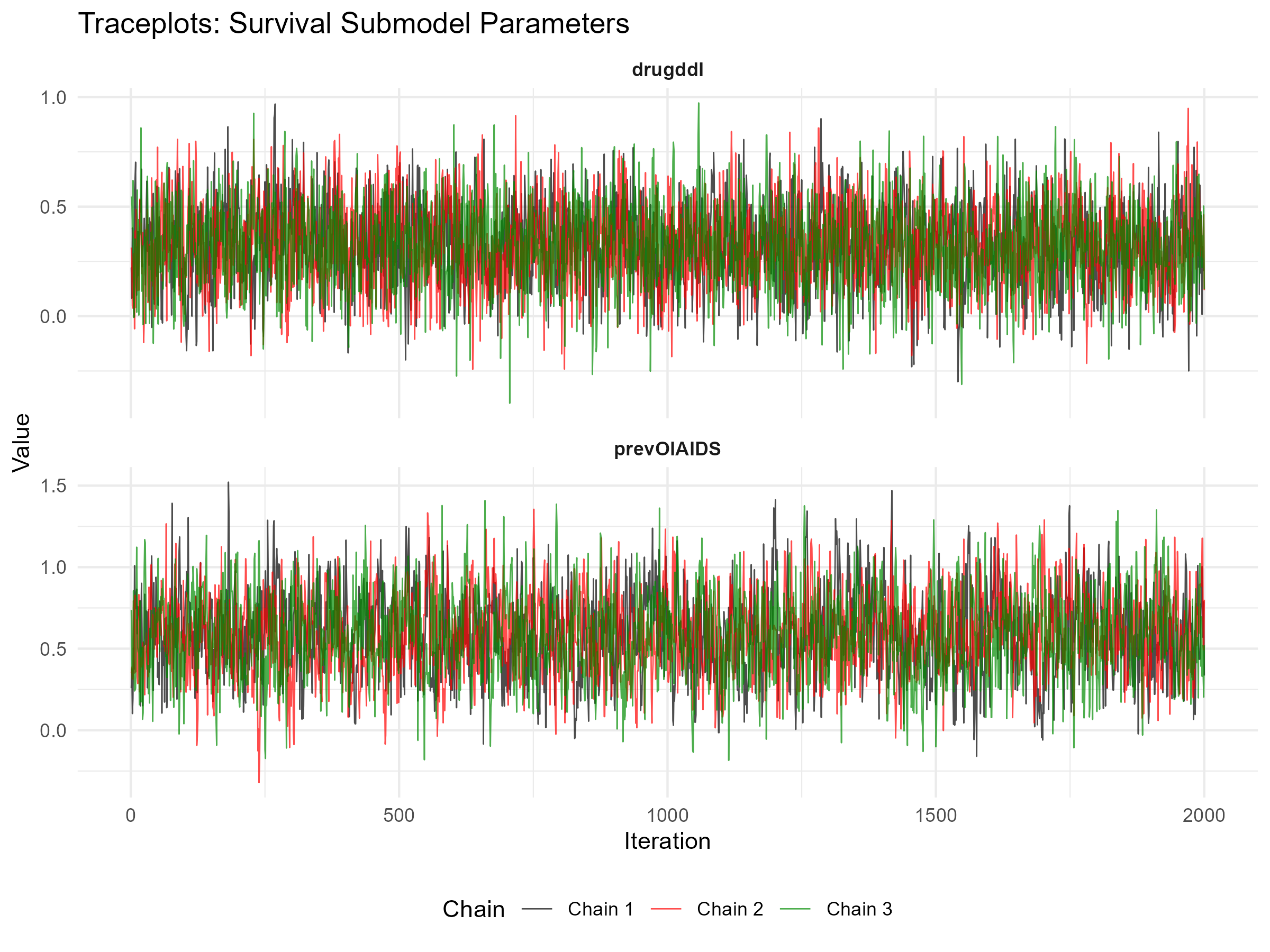}
\caption{Survival submodel parameters.}
\end{subfigure}

\caption{Traceplots for key parameters across three MCMC chains. All
monitored parameters had $\widehat{R}<1.01$.}
\label{fig:traceplots}
\end{figure}
\clearpage

\begin{figure}[H]
\ContinuedFloat
\centering
\begin{subfigure}[b]{0.85\textwidth}
\centering
\includegraphics[width=\textwidth]{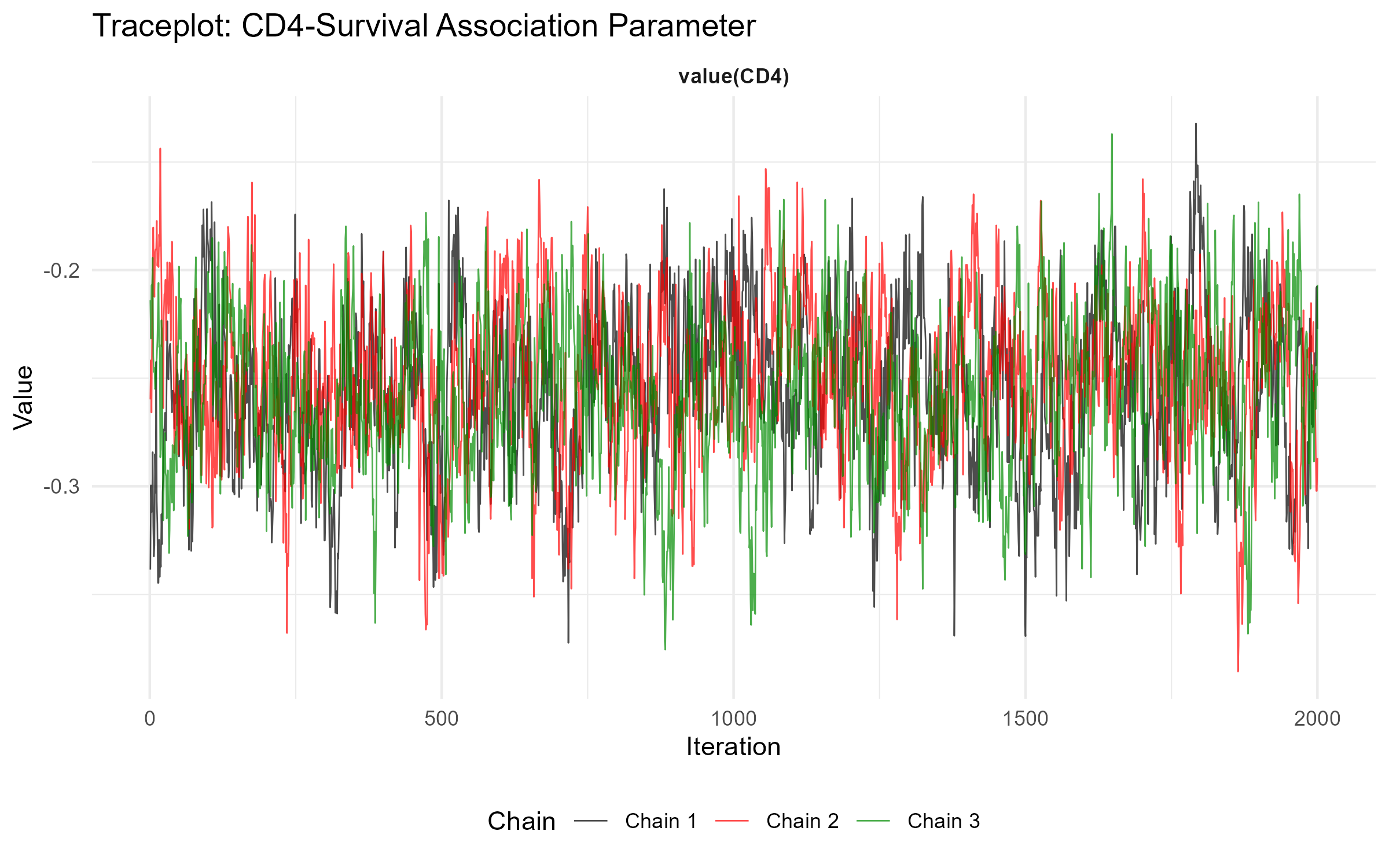}
\caption{Association parameter for the current value of CD4.}
\end{subfigure}
\caption[]{Traceplots for key parameters across three MCMC chains
(continued).}
\end{figure}

\subsection{Comparison of Methods}

\begin{figure}[H]
\centering
\includegraphics[width=0.75\textwidth]{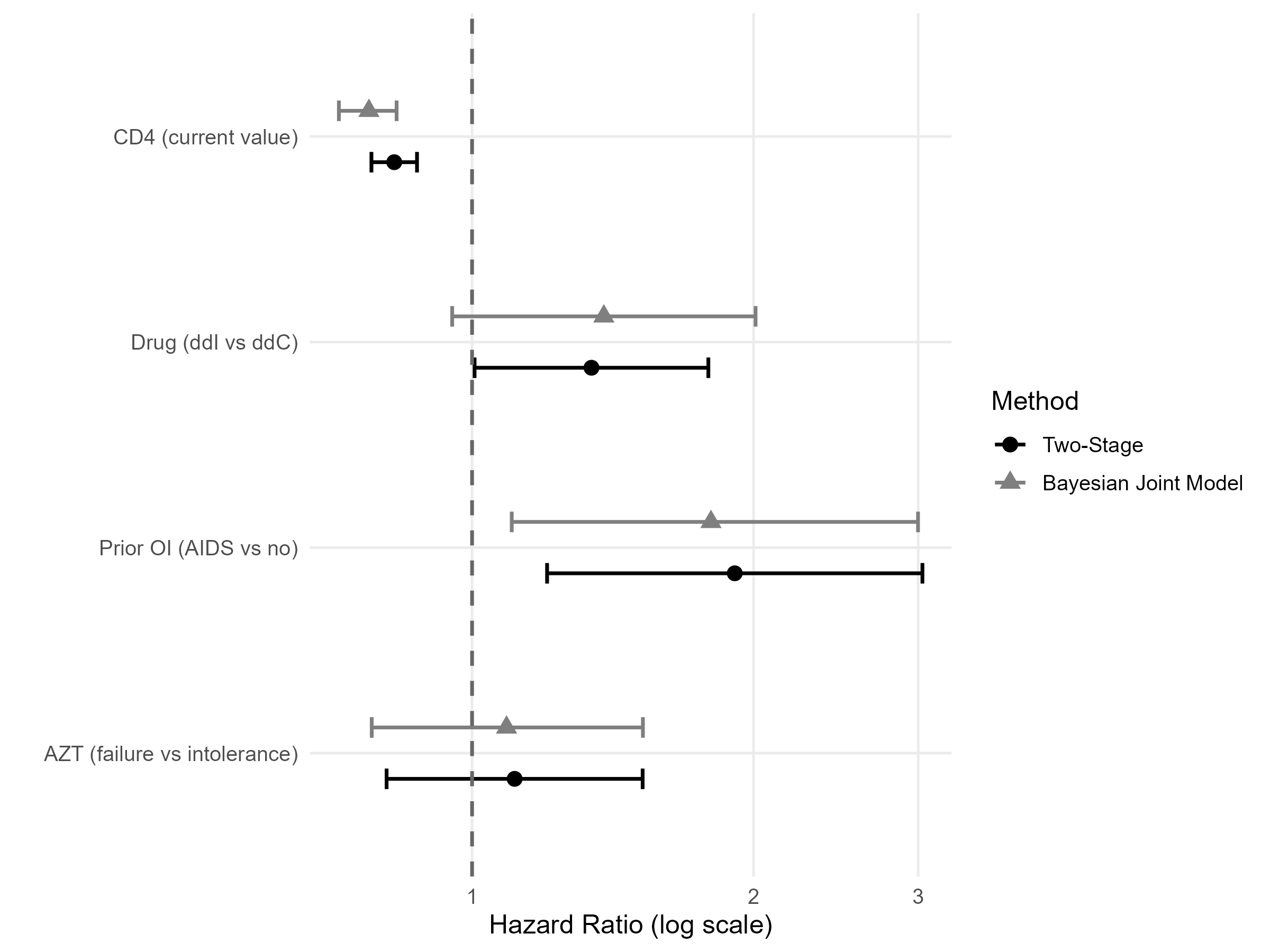}
\caption{Forest plot of hazard-ratio estimates from the baseline
two-stage approach and Bayesian joint model.}
\label{fig:forest}
\end{figure}

Table~\ref{tab:comparison} and Figure~\ref{fig:forest} compare the
estimates from the two approaches. The estimated association between
drug assignment and mortality was similar in direction and magnitude:
HR $=1.342$ under the baseline two-stage approach and HR $=1.383$ under
the joint model. The principal difference concerned uncertainty. The
two-stage confidence interval narrowly excluded 1, whereas the
joint-model credible interval included 1.

The joint model produced a stronger estimated association between CD4
and mortality. However, the CD4 hazard ratios are not directly
equivalent because the two-stage analysis used a predicted baseline CD4
value while the joint model used the current underlying CD4 value.

\begin{table}[H]
\centering
\caption{Comparison of survival-model estimates from the baseline
two-stage approach and Bayesian joint model. HR = hazard ratio;
CI = 95\% confidence interval; CrI = 95\% credible interval.}
\label{tab:comparison}
\begin{tabular}{lccccc}
\toprule
& \multicolumn{3}{c}{\textbf{Baseline Two-Stage}}
& \multicolumn{2}{c}{\textbf{Bayesian Joint Model}} \\
\cmidrule(lr){2-4}
\cmidrule(lr){5-6}
\textbf{Parameter}
& \textbf{HR}
& \textbf{95\% CI}
& \textbf{$p$}
& \textbf{HR}
& \textbf{95\% CrI} \\
\midrule
CD4, square-root scale
& 0.826 & 0.781--0.873 & $<0.001$
& 0.776 & 0.720--0.830 \\
Drug, ddI vs.\ ddC
& 1.342 & 1.006--1.789 & 0.045
& 1.383 & 0.952--2.010 \\
Prior OI, AIDS vs.\ no AIDS
& 1.910 & 1.203--3.031 & 0.006
& 1.801 & 1.103--2.999 \\
AZT, failure vs.\ intolerance
& 1.111 & 0.810--1.522 & 0.515
& 1.089 & 0.781--1.523 \\
\bottomrule
\end{tabular}

\vspace{0.25cm}
\begin{minipage}{0.75\textwidth}
\small
\textit{Note.} The two-stage CD4 estimate represents a predicted
baseline value, whereas the joint-model CD4 estimate represents the
current underlying square-root CD4 value. The two CD4 hazard ratios
therefore do not have identical interpretations. Confidence intervals
and credible intervals also arise from different inferential frameworks.
\end{minipage}
\end{table}

% ============================================================
\section{Discussion}
% ============================================================

This study compared a baseline two-stage approach with a Bayesian joint
model for repeated CD4 measurements and survival in an HIV clinical
trial. The estimated association between ddI and mortality was similar
in direction and magnitude across the two approaches, but the methods
differed in their quantification of uncertainty. The two-stage confidence
interval narrowly excluded 1, whereas the joint-model credible interval
included 1. The result should therefore not be described as a reversal
in the estimated drug association. Instead, it shows that different
modeling approaches can produce different inferential statements even
when their point estimates are similar --- which is precisely the sense
in which method choice changed statistical inference in this analysis:
the conclusion about statistical significance, not the estimated effect,
depended on the method.

The stronger CD4 association under the joint model is consistent with
established concerns about treating estimated longitudinal values as
known covariates.\cite{Tsiatis1995,Wulfsohn1997} Nevertheless, the
present comparison does not isolate the effect of measurement-error
correction because the models used different representations of CD4. The
baseline two-stage approach used a single predicted baseline value,
whereas the joint model used the evolving underlying CD4 trajectory. A
more direct comparison would use time-varying predicted CD4 values in
the second-stage Cox model.

The descriptive LOESS curves suggested possible differences in average
CD4 patterns between treatment groups. However, the longitudinal
submodel did not include a time-by-drug interaction and therefore did not
formally estimate treatment-specific slopes. The descriptive curves
should not be used to explain the difference between the survival-model
results without an additional interaction analysis.

The Bayesian implementation was computationally manageable in this
application. All three chains showed adequate convergence according to
the traceplots and Gelman--Rubin diagnostics. Joint modeling nevertheless
requires careful attention to the longitudinal and survival submodels,
the association structure, prior distributions, baseline-hazard
specification, convergence diagnostics, and interpretation.

Several limitations should be emphasized. First, the two approaches did
not use identical summaries of the longitudinal CD4 process.
Consequently, differences between their results cannot be attributed
exclusively to measurement error or informative dropout. Second, the
longitudinal submodel did not include a time-by-drug interaction, so
treatment-specific differences in CD4 slopes were not formally
evaluated. Third, the dataset was relatively small, with 467 patients,
and the findings may not generalize to other studies. Fourth,
sensitivity analyses using alternative prior specifications were not
conducted. Fifth, only a current-value association structure was
considered. Alternative structures, such as the slope of the CD4
trajectory or cumulative CD4 exposure, may yield different results.
Finally, a simulation study would provide a more controlled assessment
of when and why the methods diverge.

Despite these limitations, the analysis provides a useful applied
illustration. The way a longitudinal biomarker is summarized and linked
to a survival outcome can materially affect statistical inference.
Bayesian joint modeling offers a coherent framework for simultaneous
analysis, but its appropriateness depends on the research question,
target estimand, model assumptions, and diagnostic evidence.

% ============================================================
\section{Conclusion}
% ============================================================

The baseline two-stage approach and Bayesian joint model produced similar
point estimates for the association between drug assignment and
mortality but differed in their representation of uncertainty. The
two-stage confidence interval narrowly excluded the null, whereas the
joint-model credible interval included 1. The joint model also estimated
a stronger association between the current underlying CD4 value and
mortality.

Because the baseline two-stage analysis used a single predicted CD4
value while the joint model used the evolving underlying CD4 trajectory,
the observed differences cannot be attributed solely to measurement
error or informative dropout. Instead, the findings show that the way
longitudinal biomarker information is represented can materially affect
survival-model inference. Bayesian joint modeling provides a coherent
framework for simultaneously analyzing longitudinal and time-to-event
outcomes, but its use should be guided by the research question, target
estimand, model assumptions, and appropriate diagnostic checks. Future
work should compare the joint model with a time-varying two-stage
analysis and examine alternative association structures and prior
specifications.

% ============================================================
% DECLARATIONS
% ============================================================

\section*{Acknowledgements}
Not applicable.

\section*{Funding}
This research received no specific grant from any funding agency in the
public, commercial, or not-for-profit sectors.

\section*{Declaration of Conflicting Interests}
The author declares no conflicting interests with respect to the
research, authorship, and publication of this article.

\section*{Ethical Approval}
This study used a publicly available dataset distributed with the
\texttt{JMbayes2} R package. No participants were recruited and no
primary data were collected for this analysis.

\section*{Availability of Data and Code}
The dataset is available through the \texttt{JMbayes2} R
package.\cite{Rizopoulos2023} Analysis code is available from the
corresponding author upon reasonable request.

% ============================================================
% REFERENCES
% ============================================================


\begin{thebibliography}{99}

\bibitem{Cox1972}
Cox DR.
Regression models and life-tables.
\textit{Journal of the Royal Statistical Society: Series B}.
1972;34(2):187--220.

\bibitem{Prentice1982}
Prentice RL.
Covariate measurement errors and parameter estimation in a failure time
regression model.
\textit{Biometrika}.
1982;69(2):331--342.

\bibitem{Tsiatis1995}
Tsiatis AA, DeGruttola V, Wulfsohn MS.
Modeling the relationship of survival to longitudinal data measured
with error: applications to survival and CD4 counts in patients with AIDS.
\textit{Journal of the American Statistical Association}.
1995;90(429):27--37.

\bibitem{Wulfsohn1997}
Wulfsohn MS, Tsiatis AA.
A joint model for survival and longitudinal data measured with error.
\textit{Biometrics}.
1997;53(1):330--339.

\bibitem{Tsiatis2004}
Tsiatis AA, Davidian M.
Joint modeling of longitudinal and time-to-event data: an overview.
\textit{Statistica Sinica}.
2004;14:809--834.

\bibitem{Rizopoulos2012}
Rizopoulos D.
\textit{Joint Models for Longitudinal and Time-to-Event Data:
With Applications in R}.
Boca Raton, FL: Chapman \& Hall/CRC; 2012.

\bibitem{Goldman1996}
Goldman AI, Carlin BP, Crane LR, Launer C, Korvick JA, Deyton L,
Abrams DI.
Response of CD4 lymphocytes and clinical consequences of treatment using
ddI or ddC in patients with advanced HIV infection.
\textit{Journal of Acquired Immune Deficiency Syndromes and Human
Retrovirology}.
1996;11(2):161--169.

\bibitem{Rizopoulos2023}
Rizopoulos D.
\texttt{JMbayes2}: Extended Joint Models for Longitudinal and
Time-to-Event Data.
R package version 0.4-5; 2023.

\bibitem{RCore2025}
R Core Team.
\textit{R: A Language and Environment for Statistical Computing}.
Vienna, Austria: R Foundation for Statistical Computing; 2025.

\bibitem{Pinheiro2023}
Pinheiro J, Bates D, R Core Team.
\texttt{nlme}: Linear and Nonlinear Mixed Effects Models.
R package version 3.1-164; 2023.

\bibitem{Gelman1992}
Gelman A, Rubin DB.
Inference from iterative simulation using multiple sequences.
\textit{Statistical Science}.
1992;7(4):457--472.

\end{thebibliography}
\end{document}